# Direct observation of structural phase transformations during continuous phosphorus deposition on Cu(111)

Jiří David[1], František Jeřábek[1], Pavel Procházka[1,2], Miroslav Černý[1,2], Cristian V. Ciobanu[3], Stanislav Průša[1,2], Tomáš Šikola[1,2], Suneel Kodambaka[4], Miroslav Kolíbal[1,2,*]

[1]Institute of Physical Engineering, Brno University of Technology, Technická 2, 616 69 Brno, Czech Republic

[2]CEITEC BUT, Brno University of Technology, Purkyňova 123, 612 00 Brno, Czech Republic

[3]Department of Mechanical Engineering, Materials Science Program, Colorado School of Mines, Golden, Colorado 80401, United States

[4]Department of Materials Science and Engineering, Virginia Polytechnic Institute and State University, Blacksburg, Virginia 24061, United States

[*]kolibal.m@fme.vutbr.cz

**Abstract**

Blue phosphorene -- two-dimensional, hexagonal-structured, semiconducting phosphorus -- has gained attention as it is considered easier to synthesize on metal surfaces than its allotrope, black phosphorene. Recent studies report different structures of phosphorene, for example, on Cu(111), but the underlying mechanisms of their formation are not known. Here, using a combination of in situ ultrahigh vacuum low-energy electron microscopy and in vacuo scanning tunneling microscopy, we determine the time-evolution of the surface structure and morphology during the deposition of phosphorus on single-crystalline Cu(111). We find that during early stages of deposition, phosphorus intermixes with Cu, resulting in copper phosphide structures. With increasing surface concentration of phosphorus, the phosphide phase disappears and a blue phosphorene layer forms, followed by self-assembly of highly ordered phosphorus clusters that eventually grow into multilayer islands. We attribute the unexpected transformation of stable phosphide to a phosphorene layer, and the previously unreported multilayer growth, to the presence of a large concentration of $P_2$ dimers on the surface. Our results constitute direct evidence for a new growth mode leading to a phosphorene layer via an intermediary phase, which could underpin the growth of other 2D materials on strongly interacting substrates.

**Introduction**

The realm of two-dimensional (2D) materials is a very attractive field of research due to exceptional and often unusual properties of these materials resulting from quantum confinement in two dimensions. Large-scale fabrication of even simple elemental 2D materials can be a significant challenge as their structure and, hence, electronic properties may differ depending on the preparation technique. A typical example is phosphorene, a monolayer of phosphorus atoms, whose allotropes include black and blue phosphorene (BlackP and BlueP). The BlackP monolayer is buckled on the atomic scale and crystallizes in an orthorhombic lattice whose structural asymmetry causes an anisotropic optical response, anisotropic charge carrier mobility, etc. [1, 2] These appealing properties have driven significant efforts for the scalable fabrication of BlackP, [3, 4] which is commonly obtained by mechanical exfoliation from a bulk crystal. Blue phosphorene (BlueP) is also atomically buckled, resembles graphene due to its honeycomb structure, and exhibits properties different from BlackP.[5] It has been predicted theoretically[6,7] and with limited success experimentally observed[8-13] that phosphorus evaporated onto a metal substrate can form BlueP.

The idea of growth of 2D layers such as BlackP and BlueP on metallic single-crystals is attractive because metal substrates can be prepared on a large scale,[14] which could lead to large-area 2D phosphorene layers as well. A potential disadvantage of metallic substrates, however, is that the interaction of the electronic energy bands of the 2D material with the substrate may negatively impact the properties, prevent exfoliation,[15] or further processing of the 2D layers.

To-date, there are no reports of BlackP growth on metals. Density functional theory (DFT) simulations have predicted that BlueP layers are stable on weakly interacting substrates and that clusters of phosphorus[7,16] or metal phosphides,[17] rather than BlueP, are stable on strongly interacting metals. Experimentally, the growth of BlueP layers seems challenging even on Au(111), commonly considered to be a weakly interacting substrate. Initial reports[8,9] of BlueP formation on Au(111) were encouraging, however, later studies[18,19] argued that an Au-P alloy, instead of BlueP, was formed on Au(111). Interestingly, intercalation of Si and Te at the Au-P/Au(111) interfaces appears to yield BlueP.[10,11] Successful growths of BlueP have been reported on oxidized Cu(111)[20] and also on bare Cu(111): Kaddar et al.[12] has claimed the formation of a buckled BlueP with a lattice constant of 3.4 Å that exhibits linear dispersion at specific points of the Brillouin zone, as expected for BlueP. On the contrary, Song et al.[13] has demonstrated the growth of an ultraflat and chiral BlueP with a lattice constant of 4.1 Å on the same substrate. Both the studies conclude that further deposition of phosphorus does not lead to multilayer growth but result in a self-limited hexagonal array of 3D phosphorus islands on top of the BlueP. While the mechanisms leading to the differences in the structure (buckled vs flat) of BlueP on Cu(111) have not been identified in these two reports, a recent theoretical study[21] has suggested that the growth structure may depend on the chemical potential of the depositing phosphorus. Additionally, it is puzzling that BlueP layers are formed directly on Cu(111), a relatively strongly interacting metal compared to Au(111).

Here, we provide *direct* evidence of a hexagonal structured phosphorus monolayer *and* multilayers on Cu(111), and identify the associated growth mechanism. Using *in situ* low-energy electron microscopy and diffraction (LEEM and LEED) during the deposition of phosphorus on Cu(111) and scanning tunneling microscopy (STM), we follow the surface structural and morphological evolution as a function of time. At low phosphorus coverages, copper phosphides form on the surface. At later times, surprisingly, we observe an abrupt transformation of the phosphide into a blue phosphorene overlayer. Further deposition leads to the formation of a hexagonal array of phosphorus clusters on top of this phosphorene layer, followed by the coalescence of these clusters into 3D high aspect ratio mounds and 2D triangular islands. Our results are surprising because 1) the phosphide phase, expected to be stable on Cu(111), transforms into phosphorene and 2) multilayer islands are formed (not reported yet). We suggest (and provide justification below) that the destabilization of the phosphide occurs in presence of a large concentration of $P_2$ dimers on the surface.

**Results**

In all our experiments, phosphorus is generated predominantly as $P_2$ molecules by decomposing GaP in a Knudsen cell operated at high temperatures (~1080 K) (documented in Supporting information, Fig. S1). The rate of deposition is estimated (see Methods for details) at one monolayer (ML) of P atoms per 26 minutes on a Cu(111) substrate held at temperature $T = 450$ K (0.038 ML/min). With the increasing substrate temperature $T$ up to ~660 K, the P deposition rates decrease, presumably due to the increased desorption rate of P (and vice versa). However, the structural quality of the deposits as indicated by the LEED spot sharpness and intensity improves with increasing $T$; at $T > 660$ K, LEED shows only Cu(111)-related spots and the LEEM images reveal surface steps characteristic of bare Cu. Low energy ion scattering (LEIS) measurements has confirmed that the as-deposited surfaces are composed exclusively of phosphorus and copper (Fig. S2), i.e. the presence of impurities, if any, are below the detection limits of LEIS.

Figure 1 shows a typical sequence of bright-field LEEM images and LEED patterns acquired during the deposition of phosphorus onto the Cu(111) single-crystal held at $T = 450$ K along with schematics of the compositional and structural changes occurring at the surface. The top row in Fig. 1 shows a representative LEEM image and corresponding LEED pattern obtained from bare Cu(111). (Here $t = 0$ refers to the time at which P deposition was initiated.)

Atomically smooth terraces, monoatomic steps, and step bunches can be seen in the LEEM image and six-fold symmetric spots in the LEED pattern, all characteristics of a clean, unreconstructed Cu(111)-1×1 surface. With the onset of P deposition, the *in situ* LEEM image shows the surface covered with brighter and darker areas (domains) and LEED data reveal the emergence of new, qualitatively similar patterns as a function of $t$; one such a typical LEED pattern obtained at $t = 15$ minutes is shown in Fig. 1, second row from the top. At around $t = 26$ minutes, when the nominal amount of deposited P reaches 1 ML, the LEEM bright-field image shows a uniform surface with substrate steps visible; the initial LEED pattern disappears and new six-fold symmetric spots appear (see Fig. 1, third row from the top). Each of these spots is surrounded by six satellite spots, indicative of a higher order periodicity (a moiré) formation. After prolonged deposition, bright triangular shaped islands can be seen in the LEEM image; more spots appear in addition to the moiré pattern in the LEED (see Fig. 1, bottom row). We suggest and justify below that the deposition of phosphorus on Cu(111) results first in the formation of a transient phosphide phase, which subsequently transforms into a phosphorene monolayer, followed by the nucleation and growth of 2D and 3D islands. We will describe in detail below each of the structural transitions observed in the LEED patterns.

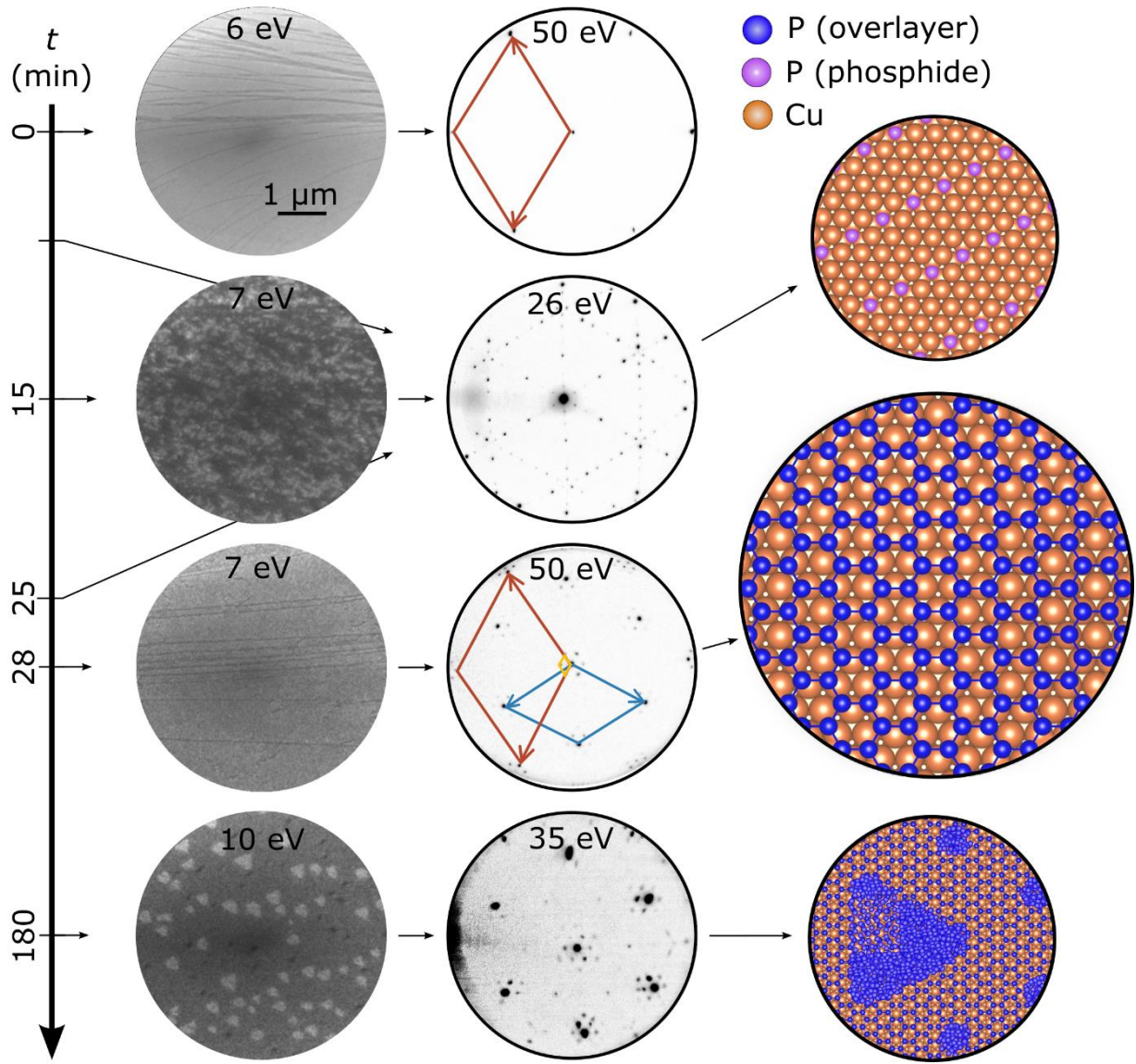

Fig. 1: Overview of the representative phase transformations observed during phosphorus deposition on Cu(111). The first and second columns are *in situ* bright-field low-energy electron microscopy (LEEM) images (5 μm field of view) and low-energy electron diffraction (LEED) patterns, respectively, acquired as a function of time $t$ during exposure to phosphorus (from a GaP effusion cell) at a deposition rate of 0.038 monolayers (ML)/min on single-crystalline Cu(111) at temperature $T = 450$ K. The third column shows schematics of the surface structures determined from the experimental data. At time $t = 0$, the Cu(111) surface is clean, after which the surface is continuously exposed to phosphorus for 180 minutes. Starting at $t \approx 15$ minutes, the Cu(111) surface is covered with a new phase, whose contrast and structure are shown in the LEEM and LEED data in the second row from the top. A detailed structural analysis of this phase is presented in Fig. 2. Despite subtle variations, the LEED pattern appears unchanged until around 25 minutes. Then, the pattern suddenly changes (within a minute) and quickly evolves into a well-defined moiré pattern, see the LEED data in the third row and Fig. 3. Further deposition results in the formation of a hexagonal array of phosphorus clusters, followed by phosphorus islands. That is accompanied by a very slow change in the diffraction pattern, as seen in the fourth row LEEM and LEED data. The electron energies for bright-field imaging are chosen to provide a distinct contrast among the different surface features: Cu(111) surface steps appear as dark lines, different phosphide superstructures, triangular and elliptical islands on top of the two-dimensional (2D) phosphorene layer appear brighter and darker grey compared to the surface.

**Initial growth phase**. First, we focus on the changes in the surface structure observed in the diffraction patterns during the early stages (i.e. $t \leq 15$ minutes) of the phosphorus deposition on bare Cu(111) at $T = 450$ K. Movie S1 in the SI shows bright-field LEEM observations during the P deposition on Cu(111). LEEM data reveal step motion and the formation of new steps on the copper surface at $t$ around 15 minutes. At initial times during the deposition, spots in the LEED patterns are barely visible, diffuse, and of weak intensities (beginning with the one in Fig. S3). Sharper, higher intensity, spots appear after approximately 15 minutes (see Figs. 1 and 2(a)). The LEED pattern shown in Fig. 2(a) is due to two superstructures: $\begin{pmatrix} 2 & 1 \\ -5 & 7 \end{pmatrix}$ and $\begin{pmatrix} 2 & 1 \\ -1 & 3 \end{pmatrix}$. The real space model of the superstructures in Fig. 2(b) shows that both are made of rows of P atoms (violet and pink). The diffraction pattern reflects the presence of three rotational domains on the substrate; areal coverages of the three domains, colored red, green, and blue, can be seen in the composite dark-field LEEM image in Fig. 2(c). Individual domains are probably even smaller than viewed by dark-field imaging (as deduced from varying intensity within individual dark-field images) and do not follow any substrate morphology. Nevertheless, the domains cover the entire surface. In order to determine the local structure of these domains, we used STM. Fig. 2(d) shows typical lower and higher magnification STM images obtained from the same sample as in Figs. 2(a) and (c) after cooling to room-temperature. The top STM image in Fig. 2(d) shows darker grey domains along three orientations (highlighted by black arrows) separated by lighter grey features. The higher magnification STM image in the bottom panel of Fig. 2(d) is acquired from a region across the interface between a domain and lighter grey region, which appears disordered (more on this later). Within the domain, periodically distributed brighter stripes can be seen. The distance between the stripes is ~0.78 nm, which is equal to the distance between the atomic rows of the $\begin{pmatrix} 2 & 1 \\ -1 & 3 \end{pmatrix}$ superstructure. The surface height profiles (not shown) of these stripes reveal picometer-scale protrusions, suggesting that the stripes are not formed by a new layer of adatoms, but rather due to rows of phosphorus atoms embedded in the Cu(111) surface.

We have also carried out density functional theory (DFT) calculations to assess the relative energetics associated with P adatoms and with intercalated P atoms. The DFT calculations show that an exchange of the copper and phosphorus atoms in the first copper layer is plausible, which could yield the surface phosphide phase. First, P adatom was traced along the path of the vector $\vec{b}$ (see Fig. 2(e)) while calculating the formation energy per phosphorus atom. Next, the phosphorus atom was inserted instead of copper in the first and the second layers. From the plot in Fig. 2(e), clearly, replacing a Cu atom in the first layer by a P atom is energetically preferred.

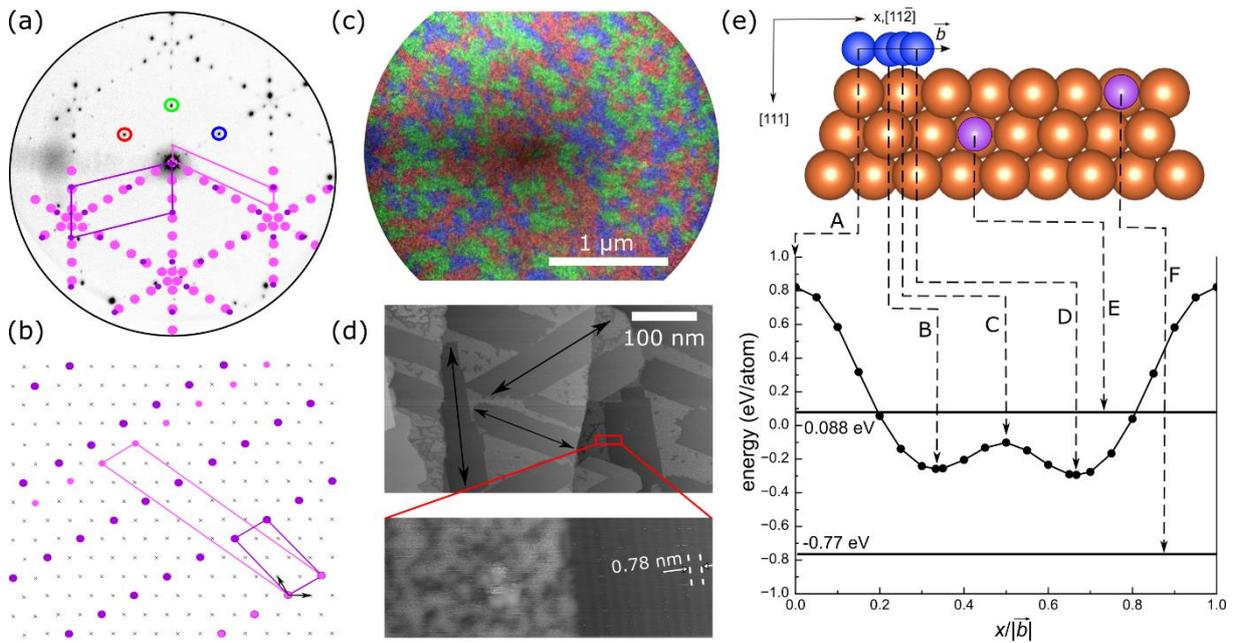

Fig. 2: Characterization of the first copper phosphide structure and density functional theory (DFT) modelling. (a) Typical LEED pattern ($E = 26$ eV) obtained from the Cu(111) sample at $T = 450$ K and $t = 15$ minutes during the P deposition at 0.038 ML/min. (b) Schematic of the atomic arrangement. In (a) and (b), violet and pink colors denote P atoms forming different superstructures on Cu(111). In (b), the first layer Cu atoms are represented with the symbol × and the primitive cell of the Cu(111) substrate is marked with black arrows. (c) Composition of dark-field LEEM images ($E = 26$ eV) corresponding to three diffraction spots in the LEED pattern in (a). Blue, red, and green colors in the image highlight three rotational domains, associated with diffraction spots marked by blue, red, and green circles, respectively, in (a). (d) Representative scanning tunneling microscopy (STM) images obtained from the same sample as in (a) after cooling to room-temperature using tunneling bias $V_T = 2.22$ V and tunneling current $I_T = 50$ pA (top) and $V_T = 2.26$ V and $I_T = 50$ pA (bottom). The black arrows indicate orientation of surface stripes seen at higher magnification. Additionally, the higher resolution bottom STM image reveals spatial periodicity of ~0.78 nm. (e) A plot showing DFT-calculated formation energy per phosphorus atom as a function of position x parallel to the vector $\vec{b} = [11\bar{2}]$ on Cu(111). Schematic showing cross-sectional view of the first three layers of Cu(111) with orange, blue, and purple colors denoting Cu, surface adsorbed P, and embedded P atoms, respectively. The lattice sites A-F are as marked in the schematic and the corresponding formation energies identified in the plot. A-D are adatoms in top (A), hollow (B,D) and bridge positions (C), E and F are phosphorus atoms embedded into the copper substrate (1st layer – F, 2nd layer – E).

**Transition to a moiré phase**. We now focus on structural changes occurring after approximately 1 ML of P deposition, which corresponds to $t = 26$ minutes in the Fig. 1 data. For $15 \leq t \leq 25$ minutes, the LEED patterns (e.g., Fig. 2(a)) vary little as more phosphide domains form. Then, within the next 60 s, the initial LEED pattern disappears and six new diffraction spots with an hexagonal arrangement emerge (see Fig. 3(a)). (The phase change is also documented *in situ* in dark-field and bright-field LEEM videos, Movies S2 and S3, respectively.) The diffraction spots are initially diffuse and elongated in both radial and azimuthal directions, suggesting the existence of rotational disorder and strain within the hexagonal layer, respectively. However, the pattern evolves quickly with the continued deposition; the spots sharpen and satellite spots appear. The latter are the first order moiré spots, a signature of two mismatched/misoriented lattices superposed on one another. Based on this observation, we conclude that the observed phase transition results in the formation of a new atomically ordered layer. It is interesting to note that the transition from the phosphide phase to a hexagonal layer occurs relatively fast (within 60 s) compared to the deposition rate (0.038 ML in 60 s) of P atoms. The number of atoms required to form a new overlayer of P, i.e. on top of the phosphide superstructure is $1.35\times10^{15}$ cm$^{-2}$ (as calculated for blue phosphorene with the lattice parameter 4.14 Å, see further), much higher than the $6.72\times10^{13}$ atoms/cm$^2$ deposited within the 60 s transition time. Furthermore, after the transition, the LEED patterns show only two sets of spots, associated to Cu(111) and the hexagonal layer. That is, there is no evidence of the presence of the previous phosphide superstructures. Therefore, we rule out the possibility of a new overlayer lying on the existing phosphide surface and suggest that the hexagonal P overlayer has been formed at the expense of the phosphide superstructure directly on Cu(111).

In order to capture the phase change with a better time resolution, we carried out another growth experiment using much lower phosphorus flux (yielding 1 ML in 6 h), stopped deposition as soon as the hexagonal pattern appeared in LEED, and cooled the sample to room-temeprature for STM characterization. The STM image in Fig. 3(b) is acquired from such a sample with ~0.78 ML of P coverage. The image shows disordered as well as highly ordered areas. The hexagonal symmetry in the region bounded by a blue square in Fig. 3(b) has been observed previously and identified as flat blue phosphorene overlayer (13). We provide supporting LEED evidence below.

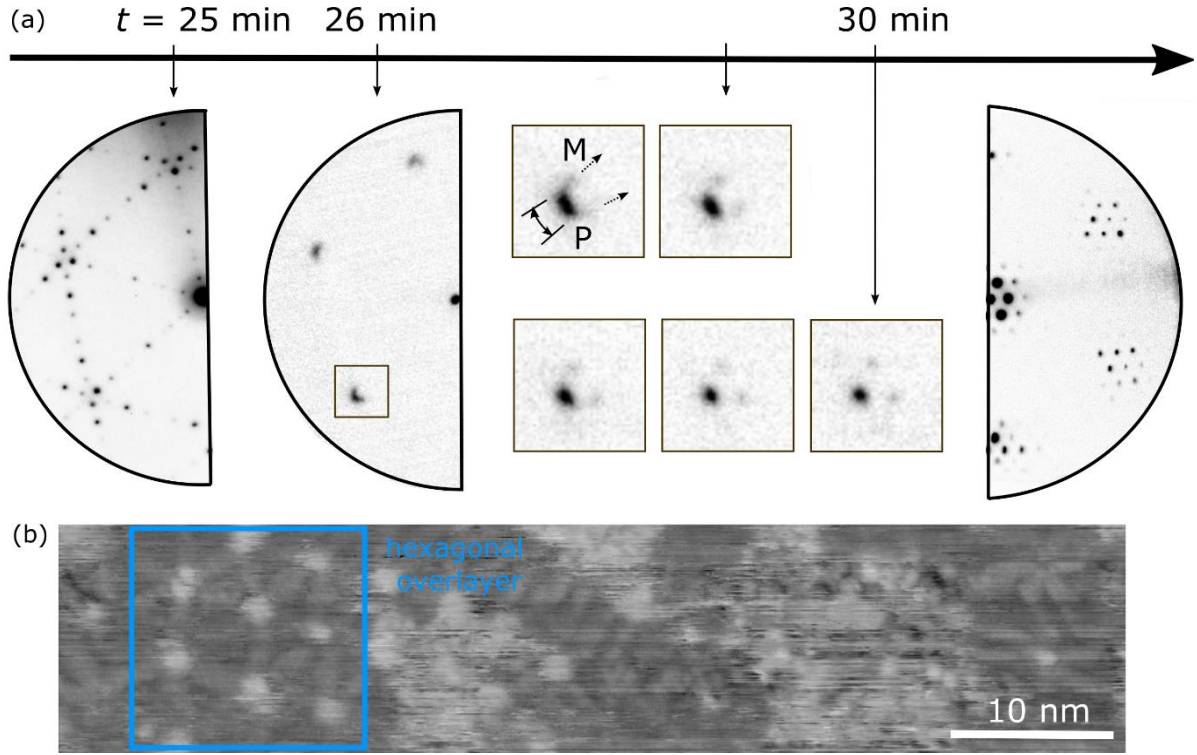

Fig. 3: (a) Phase transformations of P/Cu(111) captured in the reciprocal space. *In situ* LEED patterns ($E$ = 50 eV) from the Cu(111) sample as a function of the deposition time $t$ at $T$ = 450 K after ≈ 25 minutes of P deposition at 0.038 ML/min. Within a minute, a sudden change is observed in the LEED pattern: six-fold symmetric single, diffuse spots appear in the place of the previous multiple spot patterns due to the phosphide phase. One such a spot is labeled P in the higher magnification image of the pattern at $t$ = 26 minutes (see the inset). The spot width, a measure of its sharpness, is highlighted by the black double arrow. With increasing $t$, diffraction spots of the moiré pattern labeled M become visible, moving away from P spots as indicated by the two arrows. Concurrently, the P spots sharpen as seen in the inset at $t$ = 30 minutes. At $t$ > 30 minutes, LEED shows a sharp and stable spot pattern. Real-time dark- and bright-field LEEM images of the phase transition are shown in SI (Movies S2 and S3). (b) Representative STM image ($V_T$ = -1.6 V and $I_T$ = 53 pA) of the Cu(111) sample obtained at room-temperature after the P deposition for $t$ ≈ 390 minutes at $T$ = 450 K. The deposition rate in this experiment was ~ 0.002 ML/min.

**Moiré phase.** With the continued P deposition after the observation of the moiré pattern, the moiré spots move radially outward relative to the primary spots, suggestive of a slight change in the relative orientation of the overlayer with respect to the substrate. Finally, after approximately 30 minutes of deposition a sharp and stable moiré pattern is established as in Fig. 4(a). The emergence of the new structural phase is further confirmed by XPS data (Fig. S4), which shows that the P 2p peak shifts to higher binding energy by ~0.17 eV with the transition from the phosphide phase to the hexagonal overlayer.

Figure 4(a) shows a representative LEED pattern with higher order moiré spots, indicative of a long-range order of the emergent overlayer (we have found that the hexagonal overlayers can be grown at $T$ up to 660 K with better quality moiré patterns at higher $T$). The analysis of the spot positions in Fig. 4(a), assuming a Cu(111) in-plane lattice parameter of 2.56 Å, yields an overlayer unit cell size of 4.14 Å and a moiré periodicity of 35.8 Å. The overlayer lattice parameter is ~24% larger than that of freestanding phosphorene (3.33 Å),[6] however, similarly large lattice constants (4.10 Å in Ref. 13) and 4.20 Å in Ref. 20) have been previously reported for BlueP.

STM characterization of the same sample provides a more comprehensive picture of the surface after the phase transition to the moiré phase. Typical STM images (Figs. 4(b,c)) show a spatially periodic array of clusters with an uniform contrast on the terraces, on top of Cu(111) islands, and across the surface steps. The substrate step heights determined from STM images acquired over a wide range of tunneling biases from -3 V to + 2V are (1.93 ± 0.10) Å; this value is, within the measurement uncertainties, the same as a step height of (2.03 ± 0.04) Å obtained for pure Cu(111) prior to the P depositions, and is comparable to a step height of 2.08 Å for bulk Cu reported by others.[22] The moiré beating frequencies and hexagonal overlayer (as previously seen in Fig. 3b) are not visible in the STM image, presumably because the contrast is relatively weak compared to the brightness of the clusters which lay on top of the overlayer. Nevertheless, the observation of a continuous and conformal overlayer (deduced from the well-arranged array of clusters on top) is considered a characteristic of carpet growth of vdW layers whose orientations relative to the substrate give rise to periodic moirés with high corrugation amplitudes.[23] That is, STM results provide further (yet indirect) evidence for the formation of phosphorene, a hexagonal overlayer on Cu(111). The clusters observed in STM lay on top of phosphorene. From the Fourier transforms (such as the one in Fig. 4(c) inset) of STM images, an average spatial periodicity of the clusters is determined to be (29.0 ± 1.9) Å and the height is up to 2 Å at maximum for long deposition times (measured heights being independent of STM tunneling bias, ranging from -3 V to +2 V).

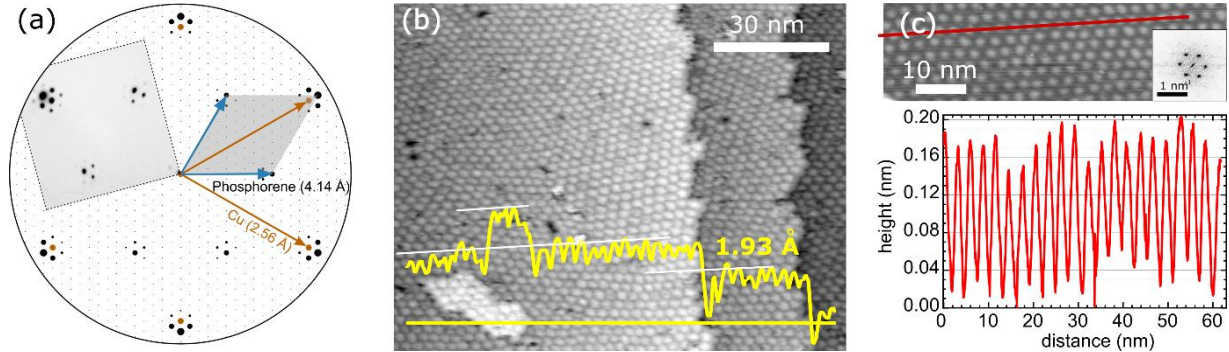

Fig. 4: Surface structure of the moiré phase. (a) Composite of the calculated diffraction pattern superposed with the experimental LEED data (the grey colored square inset) acquired using $E = 16$ eV in an off-axis geometry (sample was tilted with respect to the electrom beam) to mitigate image distortions from the Cu(111) sample at $T = 450$ K after $\approx 30$ minutes of P deposition at 0.038 ML/min. The orange and blue arrows highlight spots associated to Cu(111) and BlueP, respectively. (b,c) Typical room-temperature STM images of the same sample obtained using (b) $V_T = 2.00$ V, $I_T = 200$ pA and (c) $V_T = -2.00$ V, $I_T = 52$ pA showing highly periodic cluster island morphology, laying on top of BlueP. The yellow and red curves are surface height profiles along the yellow and red lines, respectively. The surface corrugations are periodic with ~2.8 nm peak-peak distance and 0.15 nm amplitude; hexagonal arrangement of the cluster islands is confirmed by the Fourier transform, the inset in (c), of a typical, larger STM image (not shown).

**Island phase**. Continued deposition of P at $T < 550$ K on the moiré surface results in the nucleation and growth of larger islands. Fig. 5(a) is a representative bright-field LEEM image, obtained from the same sample as in Figs. 1-4 after $t = 105$ minutes at $T = 450$ K, showing two types of islands, elliptical and triangular shaped, that appear in darker and brighter grey, respectively. The LEED pattern in Fig. 5(a) bottom panel, obtained during this measurement sequence, reveals additional diffraction spots that move with incident the electron energy $E$ (see

Movie S4 in SI), indicative of a 3D nature of the islands. (*In situ* LEED data obtained during the deposition of P at $T > 550$ K shows only moiré patterns without the extra spots, which we attribute to the absence of island growth at this *T*.) Fig. 5(b) and Movie S5 in the SI show that the islands grow and coalesce with increasing time. STM images (see Fig. S5(a)) acquired from the same sample show that elliptical islands are several nanometers tall and markedly three-dimensional. Higher resolution STM images (not shown) of elliptical islands did not reveal any atomic-scale ordering, presumably because these islands are bounded by vicinal surfaces. In comparison, the triangular islands are relatively short (Figs. 5(c) and S5(b)) with atomically-flat stepped surfaces and heights comparable (or larger) to those of the clusters in a hexagonal array (see Fig. 4(c)) and rarely over a nanometer. Interestingly, the triangular islands exhibit height variations in 0.5 Å steps (Fig. S5(c)). These steps often bunch, forming 2.0-2.5 Å „macroscopic" height step bunches (Fig. S5(d)); such atomically-corrugated surfaces are associated with unstable facets.[24] Importantly, STM images of triangular islands (see Figs. 5(c) and S5(b)) deposited at $T \sim 505$ K reveal a zigzag pattern at the island surface. However, such structure was not observed in the triangular islands grown at *T* below 473 K.

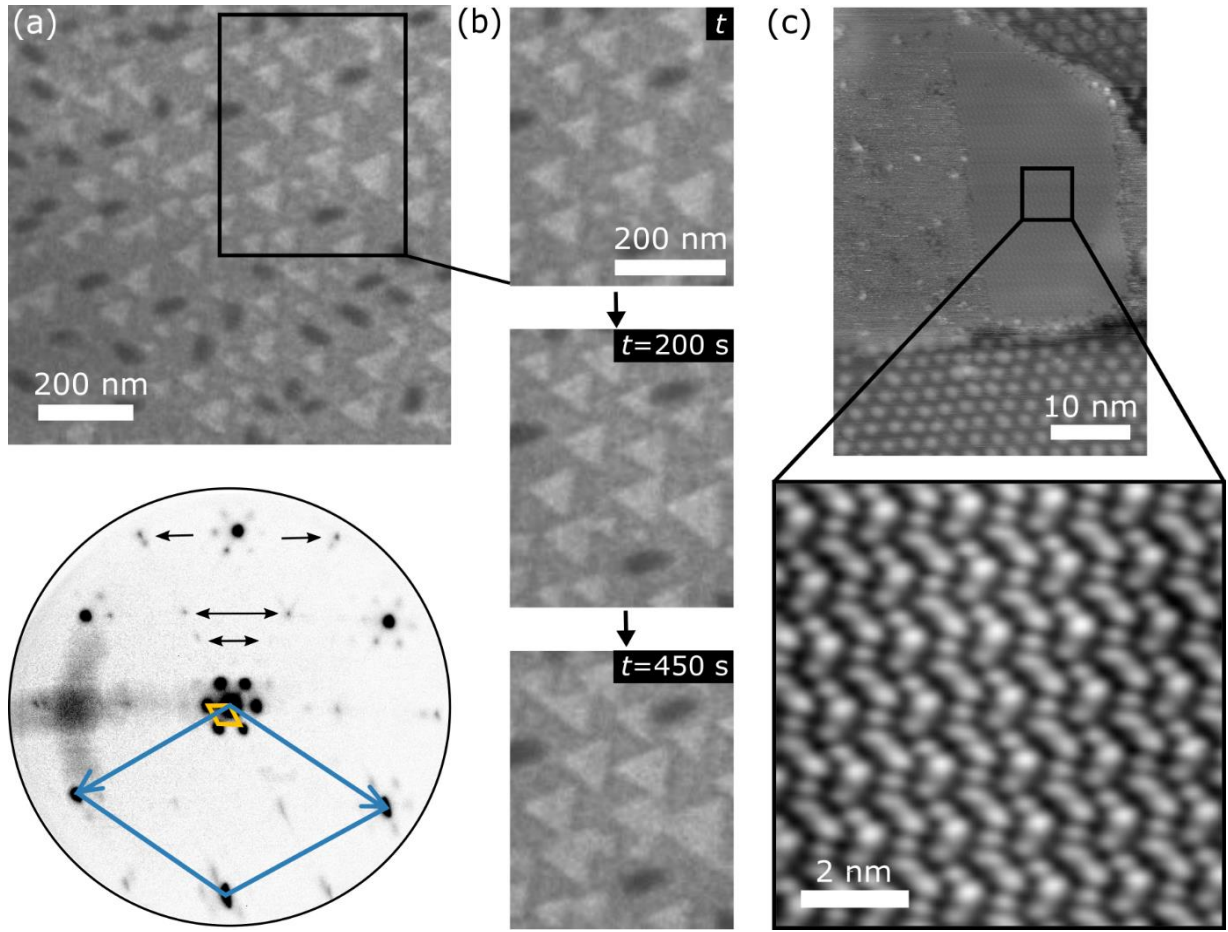

Fig. 5: Triangular and elliptical island growth and analysis. (a) Typical bright-field LEEM image ($E = 7$ eV) and LEED data ($E = 26$ eV) acquired from the Cu(111) sample at $T = 450$ K and t = 105 minutes during the P deposition at 0.038 ML/min. In the LEED pattern, the black arrows mark the spots that move with increasing $E$. (b) A series of LEEM images showing the growth of islands with increasing t during the P-deposition. Longer measurement sequences showing energy-dependent changes in LEED and morphological changes in LEEM are presented as Movie S4 and Movie S5, respectively, in SI. The electron energies are chosen to provide a distinct contrast of the substrate (medium grey), triangular islands (bright), and elliptical islands (dark). (c) Typical STM images ($V_T = 2.20$ V and $I_T = 50$ pA) of a triangular island (top) and higher magnification view (bottom) of its surface, a portion of which exhibits a zig zag pattern, obtained from a Cu(111) sample after 240 minutes of the P deposition with $T = 505$ K. The triangular island is surrounded by a hexagonal cluster array of the moiré phase.

**Discussion**

Our *in situ* LEEM and LEED data obtained during the P deposition present direct evidence for two previously unreported phenomena: 1) the growth of phosphide superstructures and subsequent transformation to BlueP hexagonal overlayers on Cu(111) and 2) the growth of phosphorus multilayer islands on top of the hexagonal overlayer. As phosphorus is expected to interact strongly with Cu(111), the formation of phosphides on Cu(111) rather than BlueP layers seems to be plausible. However, a recent theoretical study[21] has predicted otherwise and BlueP-like layers have been already experimentally demonstrated.[12,13] It is in this context that the conversion of a phosphide phase to a hexagonal phosphorene monolayer is intriguing and rather unconventional, compared to the nucleation and growth of 2D layers.[25] Furthermore, the observation of multilayer islands on top of the phosphorene layer is somewhat surprising because previous studies ruled out such a possibility. To consistently explain all of our observations, we hypothesize that there exists a critical concentration of P atoms below (above) which phosphide (phosphorene) is stable on Cu(111). It is then possible that during the P deposition, with the increasing P adatom concentration, a rapid phase change can occur resulting in phosphorene formation as proposed recently.[5] The key requirement therefore is to facilitate surface accumulation of phosphorus, for which the deposition of P monomers is preferred, due to their relatively strong bonding with Cu.[5] However, the exact nature of the depositing species depends on the phosphorus source; for example, with black phosphorus as the evaporation source, $P_4$ clusters are produced.[26] In our experiments, the decomposition of GaP results in predominantly $P_2$ dimers (see Fig. S1). DFT calculations predict that absolute value of adsorption energy of $P_2$ is higher than that of $P_4$ on Cu(111) (Fig. S6). It is also seen that on Cu(111) surface, $P_4$ clusters dissociate more easily than $P_2$ into monomers. We realize that the DFT calculations may not be applicable at the experimental conditions (e.g. elevated temperature and the presence of the P flux). Nevertheless, the DFT results could help understand the role of the phosphorus source on the differences in the resulting structures[27,28] and in controlling the growth modes of phosphorene-like layers reported by others,[12,13] being distinct to our observations. In our experiments, the P source produces predominantly stable $P_2$ dimers. It is therefore likely that the deposition flux alone may not be sufficient to provide the P supersaturation required for the phosphorene formation and hence the phosphide is the first phase being formed under these conditions. With the continued deposition, as the P concentration builds up, the phosphide phase becomes unstable and a phosphorene layer formation is favored, resulting in a rapid phase transition observed in our experiments.

The phosphide layer thus plays crucial role during deposition, serving as a reservoir of phosphorus atoms for the phase transformation, together with already present adatoms/dimers. It allows reaching very high supersaturation of phosphorus adatoms on the surface, which is not easily possible on pure Cu(111) due to the energetics of $P_2$ dimer adsorption and dissociation.[5] The kinetics of the phosphide to phosphorene phase transformation is not accounted for in a typical DFT calculation,[7] which could explain the discrepancy between prior theoretical studies and our results.

The diffraction data reveal the complex and fast transformation of the initially formed hexagonal overlayer. The radial blur of the diffraction spots results from a rotational disorder; these are the two chiral domains of phosphorene reported in Ref. 13. In agreement with Ref. 13, the further deposition causes these domains to diminish quickly, and the hexagonal overlayer flattens, finally reaching a lattice constant of 4.14 Å. Although being extremely large as compared to freestanding phosphorene, similarly large lattice constants have been reported in other works as well (4.10 Å in Ref. 13, 4.20 Å in Ref. 20). The diffraction data confirm the formation and existence of a hexagonal monolayer, as well as STM image taken during the phase transformation (Fig. 3b), being identical to the one taken by Song et al.[13]

Nevertheless, the STM images of the post-transition phase (Fig. 4c) suggest a more complex structure, where an array of clusters covers the hexagonal monolayer in a hexagonal pattern (not visible in LEED). This result has been repeatedly reported in previous papers,[12,20] and thoroughly explained in Ref. 13: the excess phosphorus atoms nucleate on top of the hexagonal overlayer and follow a self-assembly process into spatially arranged clusters.

The periodicity of the hexagonal array of phosphorus clusters seen in the STM images is 28.7 Å and significantly differs from the moiré lattice parameter (35.6 Å) deduced from the LEED data. The measured periodicity by STM is comparable to the Fermi wavelength (29 Å) of the copper surface. It has been previously reported[29] that the standing wave patterns in the electron density promote a self-assembly of Cu clusters on Cu(111). The same effect observed in our experiments could suggest that the phosphorene overlayer is ‚transparent' to the emanating fields, similar to graphene, allowing a remote epitaxy of adlayer clusters.[30] Nevertheless, this claim requires further investigation.

With the continued deposition, we observe the clusters growing in size but the growth is not self-limiting as previously reported.[12] Our data show that the extended deposition onto the cluster-covered surfaces results in the formation of low aspect ratio, quasi-2D triangular islands and relatively higher aspect ratio, 3D elliptical mounds (Fig.5). We attribute the island formation to the attachment kinetics of $P_2$ dimers, the dominant depositing species in our experiments compared to earlier reports. Finally, we comment on the highly periodic zigzag patterns observed in the STM images of the triangular islands deposite dat higher temperatures ($T \sim 505$ K). While we are not able to assign this particular pattern to BlueP, BlackP, or a moiré due to the superposition of BlueP/BlueP or BlackP/BlueP layers, clearly these islands are crystalline phosphorus structures that appear to be stable on hexagonal phosphorus layers.

**Conclusions**

Our in situ LEEM and LEED observations reveal two novel phenomena: the formation of phosphide superstructures that transition into BlueP hexagonal layers on Cu(111) and the growth of phosphorus multilayer islands atop these layers. While phosphide formation on Cu(111) is expected due to strong P-Cu interactions, our results suggest a critical P concentration beyond which phosphorene becomes stable, leading to a rapid phase transition. The initial phosphide layer likely serves as a phosphorus reservoir, facilitating the high supersaturation needed for phosphorene formation. This transition challenges previous assumptions and underscores the importance of deposition conditions and phosphorus source in controlling the growth of 2D phosphorene layers. The rapid phase transformation and subsequent island growth highlight the complex kinetics and energetics involved in this system, offering new insights into 2D material fabrication on metal substrates. The rapid phase transformation from phosphide to phosphorene suggests a possible mechanism to control the growth of 2D phosphorus layers, which could have significant implications for scalable production.

**Acknowledgements**

We would like to acknowledge Jan Čechal, Zdeněk Jakub and Matthias Blatnik for help with STM measurements and Pavel Bábík for LEIS measurements. This work was supported by the project Quantum materials for applications in sustainable technologies (QM4ST), funded as project No. CZ.02.01.01/00/22_008/0004572 by OP JAK, call Excellent Research. We acknowledge support from Brno University of Technology (specific research FSI-S-23-8336).

SK gratefully acknowledges support from the Air Force Office of Scientific Research (AFOSR, Dr. Ali Sayir) under Grant # FA9550-20-1-0184 and the National Science Foundation (NSF) for DMR Award 2245008 (old award ID 2211350) and mobility under the project

**Methods**:

All the experiments were carried out in an utrahigh vacuum (UHV) system (base pressure ~$10^{-10}$ mbar), which houses a SPECS FE-LEEM P90 low-energy electron microscope, SPECS X-ray photoelectron spectroscope (XPS) with the Mg K$\alpha$ X-Ray source and Phoibos 150 spectrometer, SPECS Aarhus 150 scanning tunneling microscope (STM) and high sensitivity IonTof Qtac100 low-energy ion spectrometer (LEIS), among other tools, and allows *in vacuo* transfer samples under UHV conditions. First, a single-crystal Cu (111) substrate was cleaned by several cycles of 2 keV Ar$^+$ bombardment at $10^{-5}$ mbar at room-temperature, followed by annealing at 785 K. Surface morphology, structure, and composition of the as prepared sample were checked by LEEM, LEED, and XPS, respectively. The substrate temperature $T$ during the phosphorus deposition ranged from 420 K to 630 K; most of the experiments were performed at 450 K. Phosphorus is generated by preferential evaporation from solid GaP chunks (MBE Komponenten, 6N purity) (for the mass spectrum of evaporated molecules, see Fig. S1), directly in the LEEM. Ga atoms are captured by a cap at the end of the crucible, which prevents contamination of Cu(111) with Ga. We measured the pressure increase in the chamber caused by evaporation (analogically to beam flux monitoring utilized in molecular beam epitaxy), which is reproducible and a stable measure of the phosphorus flux. During a typical P deposition experiment, the pressure $p$ in the LEEM chamber increased to $1.2\times10^{-9}$ mbar with the phosphorus cell at 1080 K. From STM images, we estimated the deposition rate as one monolayer of P atoms (equivalent to a surface concentration of $1.77\times 10^{15}$ P atoms on the Cu(111) surface) per 26 minutes at $T = 450$ K. It should be noted that the actual deposition rate depends on both the flux of P atoms impinging on the substrate and the sticking coefficient at a given $T$; the higher the substrate temperature the lower the deposition rate.

LEIS, XPS, and STM characterizations were carried out after passive cooling of the sample to room-temperature. The electron emission angle in XPS was fixed along the normal to the surface. Individual XPS spectra were acquired in a high magnification mode, 25 eV and 40 eV pass energies for Cu and P, respectively. All the spectra were acquired with an energy step of 0.1 eV and integrated by utilizing several sweeps. The spectra have not been shifted. The relevant photoelectron peaks were fitted with Voigt functions after a Shirley background subtraction. The scanning tunneling microscope was equipped with a KolibriSensor. STM measurements were performed at room temperature in a constant current mode; the sample bias voltages ($V_T$) and tunneling currents ($I_T$) were varied and noted with each image. The STM images have not been corrected for a drift during processing. In LEIS, the focused primary beam of He$^+$ ions impacted perpedicular to the surface over a selected area of 2 × 2 mm$^2$. The detector collected all the ions scattered at the 145° polar angle into all asimuthal angles (0 - 2π).

**Details on DFT calculations.** The total energies of all configurations were computed using the ab initio code VASP[31] employing the projector augmented-wave method[32]. The exchange and correlation contribution to the energy were evaluated with the help of the generalized gradient approximation parametrized by Perdew et al.[33]. Integrations over the Brillouin zone were performed using a mesh of equidistantly spaced k-points with a maximum distance of 0.2 Å$^{-1}$. Whenever we optimized the atomic positions, the residual forces on the atoms were relaxed below 0.01 eV/Å. The cut-off energy for the plane-wave basis was set to 500 eV. To simulate the Cu surface, we created a slab of three (111) Cu planes. The two lower planes were kept fixed and only the atoms in the upper (surface) plane were allowed to relax. In order to compare the energy of individual configurations containing $N_P$ atoms of phosphorus and $N_{Cu}$ atoms of Cu, we employed the energy of formation $E_f = \frac{E_{tot} - E_{surf} - N_P E_P - N_{Cu} E_{Cu}}{N_P}$, where $E_{tot}$, $E_{Cu}$, and $E_P$ are the energies of the particular configuration (a Cu slab with a P atom), single Cu atom in the fcc bulk and a single P atom in the blue phosphorene, respectively. $E_{surf}$ is the energy of both relaxed upper and fixed bottom surface of the Cu slab. Energy of a pure Cu slab can be then expressed as $N_P E_P + E_{surf}$. Note that if one considers adsorption of $P_n$ clusters (with n=2 for dimer and n=4 for tetramer) on the Cu surface, redefining $E_P$ to be the energy of a single P atom in the considered cluster, the definition of $E_f$ becomes equivalent to the definition of an adsoption energy (shown in Fig. S6).